# Selection of Architecture Styles using Analytic Network Process for the Optimization of Software Architecture

**Mr. K. Delhi Babu**
S.V. University, Tirupati

**Dr. P. Govinda Rajulu**
S.V. University, Tirupati

**Dr. A. Ramamohana Reddy**
S.V. University, Tirupati

**Ms. A.N. Aruna Kumari**
Sree Vidyanikethan Engg. College, Tirupati

**ABSTRACT**

*The continuing process of software systems enlargement in size and complexity becomes system design extremely important for software production. In this way, the role of software architecture is significantly important in software development. It serves as an evaluation and implementation plan for software development and software evaluation. Consequently, choosing the correct architecture is a critical issue in software engineering domain. Moreover, software architecture selection is a multicriteria decision-making problem in which different goals and objectives must be taken into consideration. In this paper, more precise and suitable decisions in selection of architecture styles have been presented by using ANP inference to support decisions of software architects in order to exploit properties of styles in the best way to optimize the design of software architecture.*

## 1. INTRODUCTION

In general the software development organizations face the problem of selecting the best design from a group of design alternatives such as various architecture styles. Architecting the systems like distributed software is a complex design activity. It involves making decisions about a number of interdependent design choices that relate to a range of design concerns. Each decision requires selecting among a number of alternatives; each of which impacts differently on various quality attributes. Additionally there are usually a number stakeholders participating in the decision making process with different, often conflicting, quality goals, and project constraints, such as cost and schedule[1]..

Consistent with the view of Mohanty (1992), the field in general has progressed from the application of linear weighting, via linear programming and integer programming, to multicriteria decision making (MCDM) models. The AHP and analytic network process (ANP) are two analytical tools for MCDM. The AHP is employed to break down large unstructured decision problems into manageable and measurable components. The ANP as the general form of AHP is powerful to deal with complex decisions where interdependence exists in a decision model. Despite the growing number of applications of AHP in various fields that involve decision-making, ANP has started to be employed in architecture style selection in software engineering fields. However, there is still a lack of papers presenting the use of ANP in typical architecture style selection from the client's perspective. This paper is intended to apply the ANP process to select a number of styles that are Plausibly undertaken.

## 2. SOFTWARE ARCHITECTURE

Software architecture represents the earliest software design decisions which are the most critical to get right and the most difficult to change down stream in the system development cycle [2]. Making use of architecture styles is one of the ways to design software systems and guarantee the satisfaction of their quality attributes [4].Selecting suitable architecture for the system and its homogeneous subsystems is an essential parameter in software systems development [3]. Many factors affect the architecture styles selection which makes it a multicriteria decision-making problem. There are some quality attributes for each architecture style that may cause different effects on different domains. In spite of having some attributes which are listed for each style in different texts, we cannot understand the extent to what advantages and disadvantages of quality and quantity attributes of architecture are considered [5].Therefore, comparing capabilities and benefits of software architectures is somehow difficult. Moreover, capabilities of software architecture styles may have been listed with respect to a special domain. It means that software architectures should be refined and completed in accordance with architects' experiments. Consequently, deciding about the architecture style(s) that should be selected depends directly on the realization/intuition of the system architect.

### 2.1 Architecture Styles

An architecture style provides us with a glossary of component and connector types and also a set of rules about the way of their combination [6]. One (or more) semantic model(s) may exist to specify the general properties of a system regarding the specifications of its elements. Therefore, the system can be described as a style or a composition of styles [7]. Styles can loosely be seen as ways to group components, where the components are related to each other through structure. It provides us with exploitation of suitable structural and evolution patterns and facilitates component, connector, and process reuse [7]. Therefore, we can take advantage of reusability of these patterns. Architecture styles classified by Shaw and Garlan in 1994 and have been applied in many projects till now.

Therefore, a set of proved qualitative and quantitative properties are defined for these reusable structures that lead to simplicity and trustworthiness in choosing and using them in new projects and special domains [7, 8]. In accordance with specified rules and frameworks of architecture styles, use of them is simple and prevents architecture users from being involved in processes and complexities of an ad-hoc architecture design. We should mention the point that some changes should be performed in architecture styles or a combination of them should be used to gain better adaptation with problem domain and increase the architecture performance.

## 2.2 Selecting Architecture Style

One of the important parts in software design process is the selection of software architecture style to design good software architecture. Choosing the most suitable architecture style(s) among existing ones can help us in satisfying functional and especially non-functional requirements correctly and precisely [8].

As we should consider different goals and objectives in architecture styles selection, such as functional and non-functional requirements, architect's priorities and domain commitments, selecting architecture styles is categorized in multi-criteria decision-making problems. The architect should select architecture styles in way that they satisfy all criteria related to the problem in the best way. Calculating the effectiveness and importance of each requirement and aggregating them are some important issues that turn the architecture style selection into a complex problem So, we need a powerful and precise decision support system to help and support system architects. For selecting architecture styles, evaluation methods and techniques [9] are usually used; but these methods do not pay attention to the abilities and capabilities of styles.

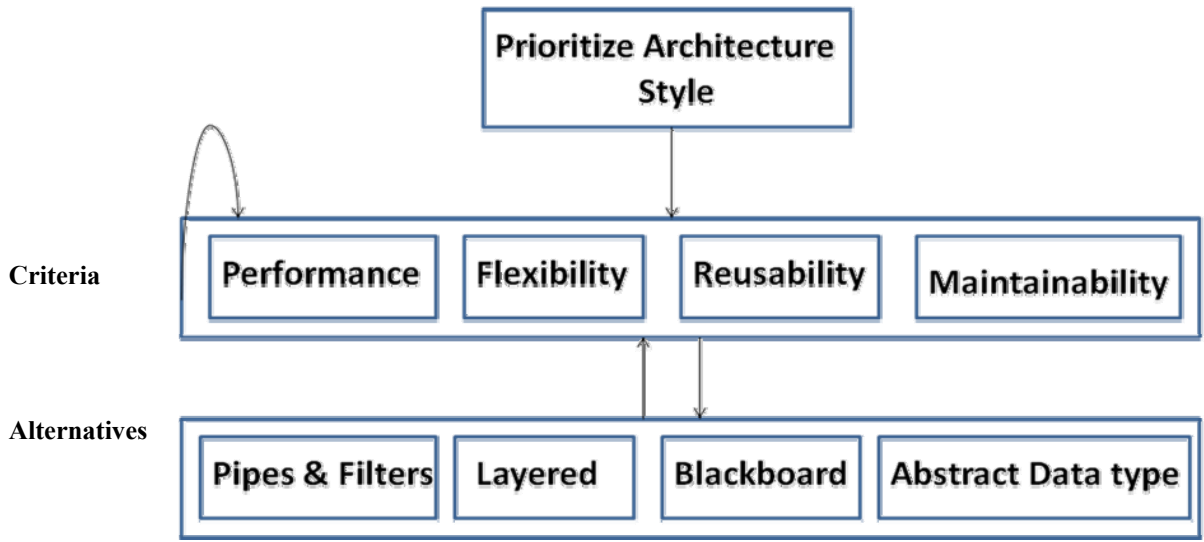

Fig 1. Frame work for architecture style selection

The above fig1 represents the frame work for architecture style selection where it contains "Criteria" cluster and "alternatives" cluster. There is a feedback between alternatives to criteria cluster which makes in comparison of criteria cluster with respect to alternatives, and also loop in criteria cluster indicates that the nodes in it are compared with themselves. These nodes in the criteria and alternatives clusters are referred from various research papers[10 11]. ANP model is best suitable to select the appropriate architecture style for the best implementation of architecture which satisfies the requirements.

## 3. ALGORITHM

The ANP incorporates both qualitative and quantitative approaches to a decision problem.

### 3.1 Qualitative approach

The four major steps for the qualitative component are described below:

1. Identify the architects problem. Suppose a client would like to select the highest scored styles from a number of potential alternative styles, the decision problem will be to "select the highest scored architecture style."

2. Ensure that the decision problem of selecting architecture style can be solved by ANP. The ANP is appropriate to solve decision problems with a network structure. Problems with a simple hierarchical model can be solved by AHP.

3. Decompose the unstructured problem of selecting the architecture style in to a set of manageable and measurable levels. The topmost level is the prioritization of styles, while the lowest level is usually the scenario or architecture styles (Saaty 1980).

4. Determine who should be responsible for making the prioritization of architecture style. Usually, various stake holders(users, domain experts) are involved for this operational management.

### 3.2 Quantitative Approach

The following describes the five major steps for the quantitative component:

1. Set up a quantitative questionnaire for collecting priorities of various stake holders (Users, Architects, Domain Experts, Programmers etc.). Saaty (1980) suggested the use of a nine-point priority scale and pair-wise comparison.

2. Estimate the relative importance between two elements criteria of selection and alternative architecture styles. Use pair-wise comparison of the elements in each matrix and calculate the eigenvector of each of the developed matrices. Refer to the existing literature having suggested the necessary algorithms for calculating the eigenvector of each matrix, such as Saaty (1980) and Cheng and Li (2001).

3. Measure the inconsistency of each of the matrices (when pairwise comparison is used) by employing the consistency ratio (CR). Refer to the existing literature having suggested the necessary algorithms to calculate CR, such as Saaty (1980) and Cheng and Li (2001). Alternatively, commercial software packages that compute eigenvectors and CRs are available (e.g., *Expert Choice for Windows*, 2003). Saaty (1994) set three acceptable levels for CR (i.e., 0.05 for 3 by 3 matrix, 0.08 for 4 by 4 matrix, and 0.1 for other matrices). Matrices that are inconsistent should be excluded or rerated by the raters.

4. Place the eigenvectors of the individual matrices (also known as sub-matrices) to form the super matrix (Saaty 1996). Refer to the later illustrative example (KWIC) of how to construct the supermatrix.

5. Ensure the supermatrix is column stochastic and raise the supermatrix to high power until the weights have been converged and remain stable (Sarkis 1999). For the purpose of mathematical computation of matrices, the authors Eddie W.L Cheng and Liang Li [12] created a program in the popular *Microsoft Excel*. Alternatively, a commercial software tool, *SuperDecisions*, developed by William J. Adams of Embry Riddle Aeronautical University and Rozann W. Saaty is appropriate to solve decision problems with a network model (Saaty 2003). Despite the availability of user-friendly software, users must have a thorough understanding of the ANP concepts before attempting to use the software. This will reduce unnecessary mistakes that hamper the making of good decisions of selecting best architecture style. In a later section, we reveal a paper exhibiting wrong results from using ANP.

## 4. ANP MODEL FOR SOFTWARE ARCHITECTURE STYLE SELECTION

Details on the Analytic Network Process (ANP) can be found in Saaty (1996), however, the main steps are summarized here for completeness.

(i) Pairwise comparisons on the elements and relative weight estimation

The determination of relative weights in ANP is based on the pairwise comparison of the elements in each level. These pairwise comparisons are conducted with respect to their relative importance towards their control criterion based on the principle of AHP and measured using Saaty's 1-to-9 scale (see table 1). The score of aij in the pairwise comparison matrix represents the relative importance of the element on row (i) over the element on column (j), i.e., $a_{ij}=w_i/w_j$

**Table 1.** The Fundamental Scale for Making Judgments

| 1 | Equal |
|---|---|
| 2 | Between Equal and Moderate |
| 3 | Moderate |
| 4 | Between Moderate and Strong |
| 5 | Strong |
| 6 | Between Strong and Very Strong |
| 7 | Very Strong |
| 8 | Between Very Strong and Extreme |
| 9 | Extreme |

With respect to any criterion, pairwise comparisons are performed in two levels, i.e. the element level and the cluster level comparison.

If there are n elements to be compared, the comparison matrix A is defined as:

$$A = \begin{pmatrix} w_1/w_1 & w_1/w_2 & \cdots & w_1/w_n \\ w_2/w_1 & w_2/w_2 & \cdots & w_2/w_n \\ \cdots & \cdots & \cdots & \cdots \\ w_n/w_1 & w_n/w_2 & \cdots & w_n/w_n \end{pmatrix} = \begin{bmatrix} 1 & a_{12} & \cdots & a_{1n} \\ a_{21} & 1 & \cdots & a_{2n} \\ \cdots & \cdots & \cdots & \cdots \\ a_{n1} & a_{n2} & \cdots & 1 \end{bmatrix}$$

After all pairwise comparisons are completed the priority weight vector (w) is computed as the unique solution of

$$Aw = \lambda_{max} w$$

Where $\lambda_{max}$ is the largest eigenvalue of matrix A and w is its eigenvector.

The consistency index [CI] and consistency ratio [CR] of the pairwise comparison matrix could then be calculated by:

$$CI = \frac{\lambda_{max} - n}{n-1}, \quad CR = CI/RCI$$

Where n is the order of comparison matrix.

RCI being a Random Consistency Index provided by Saaty (1980)

In general, if CI is less than 0.1, the judgment can be considered as consistent.

(ii) Construction of the original supermatrix (unweighted supermatrix).

The resulting relative importance weights (eigenvectors) in pairwise comparison matrices are placed within a supermatrix that represents the interrelationships of all elements in the system. The general structure of the supermatrix is described in table 3, where Ci denotes the ith cluster, eji denotes the jth element of the ith cluster and Wik is a block matrix consisting of priority weight vectors of the influence of the element in the ith cluster with respect to the kth cluster

**Table 2:** General structure of super matrix

|   |     | $C_1$ |  |  | $C_2$ |  |  |  | $C_n$ |  |  |
|---|---|---|---|---|---|---|---|---|---|---|---|
|   |     | $e_{11}$ $e_{12}$ $\cdots$ $e_{1n}$ | | | $e_{21}$ $e_{22}$ $\cdots$ $e_{2n}$ | | | | $e_{n1}$ $e_{n2}$ $\cdots$ $e_{nn}$ | | |
| $C_1$ | $e_{11}$ $e_{12}$ $\cdots$ $e_{1n}$ | $W_{11}$ | | | $W_{12}$ | | | | $W_{1n}$ | | |
| $C_2$ | $e_{21}$ $e_{22}$ $\cdots$ $e_{2n}$ | $W_{21}$ | | | $W_{22}$ | | | | $W_{2n}$ | | |
| $C_n$ | $e_{n1}$ $e_{n2}$ $\cdots$ $e_{nn}$ | $W_{n1}$ | | | $W_{n2}$ | | | | $W_{nn}$ | | |

(iii) Constructing the weighted supermatrix

The following step consists of the weighting of the blocks of the unweighted supermatrix, by the corresponding priorities of the clusters, so that it can be column stochastic (weighted supermatrix). The weighting of the clusters has to be conducted again by means of standard AHP.

(iv) Calculation of the global priority weights.

Raising the weighted supermatrix to limiting powers until the weights converge and remain stable the limit supermatrix will be obtained. In this matrix, the elements of each column represent the final weights of the different elements considered.

**5. CASE STUDY**

In order to clarify the way of making use of ANP model, we take advantage of the famous KWIC [10] example and show how to select the best architecture style. The Key Word In Context (KWIC) problem was first introduced by David Parnas and used to contrast different criteria for decomposing a system into modules [4]. The problem is defined as bellow [4]:

The KWIC system index system accepts an ordered set of lines, each line is an ordered set of words, and each word is an ordered set of characters. Any line may be "circularly shifted" by repeatedly removing the first word and appending it at the end of the line. The KWIC index system outputs a listing of all circular shifts of all lines in alphabetical order. KWIC has been widely used in Computer Science till now; for example in Unix Man page permutated index and in libraries. Different criteria could be considered to compare different architecture styles and show their differences in satisfying requirements of the same problem. The comparison criteria could be performance, reuse, change in algorithm, change in data representation, and change in function [4]. These criteria are used to compare five styles which are pipes-and-filters, layered, blackboard, abstract data type.

We have four criteria named performance (P), flexibility (F), and maintenance (M),reusability (R). There are four styles in the styles repository: pipes-and-filters, layered, blackboard, Abstract datatype. The level of satisfying each quality attribute by means of different styles has been represented in table 3.

According to our online survey the level of importance of the quality attributes via performance, flexibility, reusability and maintenance in the project at hand is 4, 3, 2 and 5 respectively.

The authors of this paper have developed code for this case study [KWIC] [12] in Java.

**Table 3:** Level of satisfying each quality attribute

| Architecture-Styles | Performance | Flexibility | Reusability | Maintainability |
|---|---|---|---|---|
| Pipes & Filters | 15 | 14 | 12 | 8 |
| Layered | 3 | 9 | 10 | 11 |
| Blackboard | 8 | 11 | 8 | 8 |
| Abstract Data Type | 13 | 7 | 10 | 11 |

The Table 3 shows relative weights of architecture styles and attributes obtained from online survey.

**As per the algorithm various comparison matrices are computed and given as follows:**

1. With respect to Prioritize:

|   | P | F | R | M | E.V |
|---|---|---|---|---|---|
| P | 1 | 2 | 3 | 1/2 | 0.277 |
| F | 1/2 | 1 | 2 | 1/3 | 0.161 |
| R | 1/3 | 1/2 | 1 | 1/4 | 0.096 |
| M | 2 | 3 | 4 | 1 | 0.466 |
|   |   |   |   |   | CR=0.0006 |

2. With respect to Performance

|   | F | R | M | E.V |
|---|---|---|---|---|
| F | 1 | 2 | 1/3 | 0.239 |
| R | 1/2 | 1 | 1/4 | 0.137 |
| M | 3 | 4 | 1 | 0.623 |
|   |   |   |   | CR =0.016 |

3. With respect to Flexibility

|   | P | R | M | E.V |
|---|---|---|---|---|
| P | 1 | 3 | 1/2 | 0.320 |
| R | 1/3 | 1 | 1/4 | 0.123 |
| M | 2 | 4 | 1 | 0.557 |
|   |   |   |   | CR=0.020 |

4. With respect to Reusability

|   | P | F | M | E.V |
|---|---|---|---|---|
| P | 1 | 2 | 1/2 | 0.297 |
| F | 1/2 | 1 | 1/3 | 0.163 |
| M | 2 | 3 | 1 | 0.539 |
|   |   |   |   | CR=0.0033 |

5. With respect to Maintenance

|   | P | F | R | E.V |
|---|---|---|---|---|
| P | 1 | 2 | 3 | 0.539 |
| F | 1/2 | 1 | 2 | 0.297 |
| R | 1/3 | 1/2 | 1 | 0.164 |
|   |   |   |   | CR =0.0090 |

6. With respect to Performance

|   | PF | L | BB | ADT | E.V |
|---|---|---|---|---|---|
| PF | 1 | 9 | 8 | 3 | 0.557 |
| L | 1/9 | 1 | 1/6 | 1/9 | 0.036 |
| BB | 1/8 | 6 | 1 | 1/6 | 0.106 |
| ADT | 1/3 | 9 | 6 | 1 | 0.300 |
|   |   |   |   |   | CR=0.245 |

7. With respect to Flexibility

|   | PF | L | BB | ADT | E.V |
|---|---|---|---|---|---|
| PF | 1 | 6 | 4 | 8 | 0.590 |
| L | 1/6 | 1 | 1/3 | 3 | 0.117 |
| BB | 1/4 | 3 | 1 | 5 | 0.238 |
| ADT | 1/8 | 1/3 | 1/5 | 1 | 0.052 |
|   |   |   |   |   | CR=0.081 |

8. With respect to Reusability

|   | PF | L | BB | ADT | E.V |
|---|---|---|---|---|---|
| PF | 1 | 3 | 5 | 3 | 0.519 |
| L | 1/3 | 1 | 3 | 1 | 0.201 |
| BB | 1/5 | 1/3 | 1 | 1/3 | 0.079 |
| ADT | 1/3 | 1 | 3 | 1 | 0.200 |
|   |   |   |   |   | CR=0.020 |

9. With respect to Maintenance

|   | PF | L | BB | ADT | E.V |
|---|---|---|---|---|---|
| PF | 1 | 1/4 | 1 | 1/5 | 0.089 |
| L | 4 | 1 | 4 | 1/2 | 0.319 |
| BB | 1 | 1/4 | 1 | 1/5 | 0.089 |
| ADT | 5 | 2 | 5 | 1 | 0.501 |
|   |   |   |   |   | CR=0.009 |

10. With respect to Pipes & Filters

|   | P | F | R | M | E.V |
|---|---|---|---|---|---|
| P | 1 | 2 | 4 | 8 | 0.466 |
| F | 1/2 | 1 | 3 | 7 | 0.320 |
| R | 1/4 | 1/3 | 1 | 5 | 0.157 |
| M | 1/8 | 1/7 | 1/5 | 1 | 0.041 |
|   |   |   |   |   | CR=0.020 |

11. With respect to Layered

|   | P | F | R | M | E.V |
|---|---|---|---|---|---|
| P | 1 | 1/7 | 1/8 | 1/9 | 0.038 |
| F | 7 | 1 | 1/2 | 1/3 | 0.188 |
| R | 8 | 2 | 1 | 1/2 | 0.294 |
| M | 9 | 3 | 2 | 1 | 0.478 |
|   |   |   |   |   | CR=0.0429 |

12. With respect to Black Board

|   | P | F | R | M | E.V |
|---|---|---|---|---|---|
| P | 1 | 1/4 | 1 | 1 | 0.143 |
| F | 4 | 1 | 4 | 4 | 0.571 |
| R | 1 | 1/4 | 1 | 1 | 0.143 |
| M | 1 | 1/4 | 1 | 1 | 0.443 |
|   |   |   |   |   | CR=0.008 |

13. With respect to Abstract Data Type

|   | P | F | R | M | E.V |
|---|---|---|---|---|---|
| P | 1 | 7 | 4 | 2 | 0.493 |
| F | 1/7 | 1 | 1/4 | 1/6 | 0.052 |
| R | 1/4 | 4 | 1 | 1/3 | 0.142 |
| M | 1/2 | 6 | 3 | 1 | 0.311 |
|   |   |   |   |   | CR=0.0488 |

These comparison matrices are used to construct an unweighted super matrix using algorithm as :

| Cluster Node labels | | 1.Prioritize | 2. Criteria | | | | 3.Alternatives | | | |
|---|---|---|---|---|---|---|---|---|---|---|
| | | Prioritize | 1.P | 2.F | 3.R | 4.M | 1.PF | 2.L | 3.BB | 4.ADT |
| 1.Prioritize | Prioritize | 0.0000 | 0.0000 | 0.0000 | 0.0000 | 0.0000 | 0.0000 | 0.0000 | 0.0000 | 0.0000 |
| 2. Criteria | 1.P | 0.2771 | 0.0000 | 0.3196 | 0.2969 | 0.5396 | 0.4735 | 0.0378 | 0.1428 | 0.4964 |
| | 2.F | 0.1600 | 0.2385 | 0.0000 | 0.1634 | 0.2969 | 0.3259 | 0.1853 | 0.5714 | 0.0509 |
| | 3.R | 0.0954 | 0.1365 | 0.1219 | 0.0000 | 0.1634 | 0.1564 | 0.2956 | 0.1428 | 0.1393 |
| | 4.M | 0.4672 | 0.6250 | 0.5584 | 0.5396 | 0.0000 | 0.0440 | 0.4812 | 0.1428 | 0.3132 |
| 3.Alternatives | 1.PF | 0.0000 | 0.5696 | 0.6034 | 0.5222 | 0.0889 | 0.0000 | 0.0000 | 0.0000 | 0.0000 |
| | 2.L | 0.0000 | 0.0328 | 0.1114 | 0.1998 | 0.3182 | 0.0000 | 0.0000 | 0.0000 | 0.0000 |
| | 3.BB | 0.0000 | 0.0930 | 0.2344 | 0.0780 | 0.0889 | 0.0000 | 0.0000 | 0.0000 | 0.0000 |
| | 4.ADT | 0.0000 | 0.3044 | 0.0506 | 0.1998 | 0.5039 | 0.0000 | 0.0000 | 0.0000 | 0.0000 |

Limit matrix is obtained as

| Cluster Node labels | | 1.Prioritize | 2. Criteria | | | | 3.Alternatives | | | |
|---|---|---|---|---|---|---|---|---|---|---|
| | | Prioritize | 1.P | 2.F | 3.R | 4.M | 1.PF | 2.L | 3.BB | 4.ADT |
| 1.Prioritize | Prioritize | 0.0000 | 0.0000 | 0.0000 | 0.0000 | 0.0000 | 0.0000 | 0.0000 | 0.0000 | 0.0000 |
| 2. Criteria | 1.P | 0.2574 | 0.2574 | 0.2574 | 0.2574 | 0.2574 | 0.2574 | 0.2574 | 0.2574 | 0.2574 |
| | 2.F | 0.1738 | 0.1738 | 0.1738 | 0.1738 | 0.1738 | 0.1738 | 0.1738 | 0.1738 | 0.1738 |
| | 3.R | 0.1119 | 0.1119 | 0.1119 | 0.1119 | 0.1119 | 0.1119 | 0.1119 | 0.1119 | 0.1119 |
| | 4.M | 0.2902 | 0.2902 | 0.2902 | 0.2902 | 0.2902 | 0.2902 | 0.2902 | 0.2902 | 0.2902 |
| 3.Alternatives | 1.PF | 0.0671 | 0.0671 | 0.0671 | 0.0671 | 0.0671 | 0.0671 | 0.0671 | 0.0671 | 0.0671 |
| | 2.L | 0.0285 | 0.0285 | 0.0285 | 0.0285 | 0.0285 | 0.0285 | 0.0285 | 0.0285 | 0.0285 |
| | 3.BB | 0.0198 | 0.0198 | 0.0198 | 0.0198 | 0.0198 | 0.0198 | 0.0198 | 0.0198 | 0.0198 |
| | 4.ADT | 0.0511 | 0.0511 | 0.0511 | 0.0511 | 0.0511 | 0.0511 | 0.0511 | 0.0511 | 0.0511 |

| P | Performance |
|---|---|
| F | Flexibility |
| R | Re usability |
| M | Maintenance |
| PF | Pipes & Filters |
| L | Layered |
| BB | Black Board |
| ADT | Abstract Data Type |

**Table 4** The meaning of the abbreviations

| Pipes & Filters | 0.0671 |
|---|---|
| Layered | 0.0285 |
| Black Board | 0.0198 |
| Abstract Data Type | 0.0511 |

**Table 5** Weights of alternatives from limit matrix.

From the limit matrix it is cleared that "PIPES & FILTERS" among the alternatives has got highest weight i.e. 0.0671, so it is selected.

**6. CONCLUSION:**

Key features of architectural styles are different with each other and, therefore, each one has its own strengths and weaknesses in a common problem. Furthermore, quality attributes which are satisfied by the same architecture interact with each other.

Consequently, identifying the strengths and weaknesses and considering criteria interaction when selecting the architecture for the problem could help architects to make more precise decisions. Architecture style selection is the essential phase in designing software systems because satisfying quality attributes is one of the important issues in designing systems that suitable architecture can fulfill them. In this way, the system architecture is propounded as a key element that we can guarantee the quality of our product by making use of that.

In this paper, ANP approach is used to represent the various criteria, and concepts of quality attributes more precisely and efficiently was defined. The Frame work designed in this paper is used to include various criteria in criteria cluster and alternatives in alternatives cluster based on our requirements, it uses feedback, loops and also considers weights from various stake holders(users, domain experts, survey etc) to select the best alternative in an efficient manner. The results obtained from this approach are better than the any other approaches followed to select architecture style such are decision support systems, Fuzzy approach, AHP, AHP-GP.

Future approach for ANP is ANP-GP, which uses Goal Programming in addition to ANP.

## 7. References

[1] Chung L et al, "Non-Functional Requirements in Software Engineering": Kluwer Academic Publishers, Boston, MA. 1999.

[2] Bass, L., Clements, P., Kazman, R., "Software Architecture in Practice", Addison-Wesley Professional, 2nd edition, 2003.

[3] Shaw, M., Clements, P., "The Golden Age of Software Architecture", IEEE, March/April, 2006.

[4] Shaw, M., Garlan, D., "Software Architecture: Perspectives on an Emerging Discipline", Prentice Hall, 1996.

[5] Svahnberg, M., "Supporting Software Architecture Evolution - Architecture Selection and Variability", Ph.D. Thesis, Blekinge Institute of Technology, Dissertation Series no. 2003:03, 2003

[6] Shaw, M., Garlan, D., *"Software Architecture: Perspectives on an Emerging Discipline"*, Prentice Hall, 1996.

[7] Bass, L., Clements, P., Kazman, R., *"Software Architecture in Practice"*, Addison-Wesley Professional, 2nd edition, 2003

[8] Albin, Stephen T., *"The Art of Software Architecture: Design Methods and Techniques",* Wiley, 1st edition, 2003..

[9] Dobrica, L., Niemela, E., "*A Survey on Software Architecture Analysis Methods"*, IEEE Transaction on software engineering, 2005.

[10] Shahrouz Moaven, Jafar Habibi, Hamed Ahmadi, Ali Kamandi "*A Fuzzy Model for Solving Architecture Styles Selection Multi-Criteria Problem*" IEEE, 2008.

[11]. Shahrouz Moaven, Jafar Habibi, Hamed Ahmadi, Ali Kamandi "*A Decision Support System for Software Architecture-Style Selection*" IEEE, 2008.

[12] ANP applied to project selection "Journal of construction engineering and management", April, 2005.

Authors:

1. **Mr. K. Delhi Babu**
   S.V. University, Tirupati
   kdb_babu@yahoo.com

2. **Dr. P. Govinda Rajulu**
   S.V. University, Tirupati
   Pgovindarajulu@yahoo.com

3. **Dr. A. Ramamohana Reddy**
   S.V. University, Tirupati

   aramamohanasvu1@yahoo.com

4. **Ms. A.N. Aruna Kumari**
   Sree Vidyanikethan Engg. College

   Tirupati

   kolla_aruna@yahoo.com